\documentclass[amsmath,amssymb,aps,reprint,superscriptaddress]{revtex4-1}
\usepackage{graphicx}
\usepackage{dcolumn}
\usepackage{bm}

\newcommand\p{\partial}

\newcommand\ep{\varepsilon}

\newcommand\di{\textrm{d}}
\newcommand\ex{\textrm{e}}

\begin{document}

\preprint{APS/123-QED}

\title{ Longwave nonlinear theory for chemically active droplet division instability }

\author{Mohammad Abu Hamed}
\affiliation{Department of Mathematics, Technion-Israel Institute of Technology, Haifa 32000, Israel }
\affiliation{Department of Mathematics, The College of Sakhnin - Academic College for Teacher Education, Sakhnin 30810, Israel }

\author{Alexander A. Nepomnyashchy}
\affiliation{Department of Mathematics, Technion-Israel Institute of Technology, Haifa 32000, Israel }

\begin{abstract}
It has been suggested recently that growth and division of a protocell could be modeled by a chemically active droplet with simple chemical reactions driven by an external fuel supply. This model is called the continuum model. Indeed it's numerical simulation reveals a shape instability which results in droplet division into two smaller droplets of equal size resembling cell division  \cite{David-Rabea2017}.
In this paper, we investigate the reduced version of the continuum model, which is called the effective model. This model is studied both in the linear and nonlinear regime. First, we perform a linear stability analysis for the flat interface, and then we develop a nonlinear theory using the longwave approach. We find that the interface at the leading order is governed by the modified Kuramoto-Sivashinsky equation. Therefore the interface is subject to a logarithmic blow up after a finite time. In addition, an expression for the interface local velocity is obtained.
\end{abstract}

\maketitle

\section{INTRODUCTION}

Over the past years, the dynamics and stability of active droplets
undergoing chemical reactions is a subject of extensive
studies \cite{Lach2016}. In addition to numerous technological
applications, active droplets provide a model of a biological cell.
Specifically, they exhibit self-propulsion \cite{Seemann2016}, growth and
spontaneous division \cite{MaciaSole2007, BrowneWalker2010, Patashinski2012, GiomiDeSimone2014}.

Recently, a simple model of the active droplet division has been suggested
in \cite{David-Rabea2017}, which includes only two components $A$ and $B$.
The droplet material $B$, which is surrounded by a solvent, is subject to
a spontaneous chemichal reaction
\begin{equation}\label{cr1}
B\rightarrow A.
\end{equation}
 Molecules $A$ are soluble, hence they leave the droplet and move to the
solvent, where they induce the chemical reaction,
\begin{equation}\label{cr2}
A+C \rightarrow B + C';
\end{equation}
here $C$ is the fuel and $C'$ is the product. Finally, material $B$
diffuses inside the droplet thus completing the reaction cycle, see Fig \ref{division}.

The problem was considered using two models,
(i) the continuum model with a diffuse interface between phases and a
continuous reaction function; (ii) the effective model with a sharp
interface and piecewise linear reaction function. In the numerical
simulations carried out for a spherical droplet, depending on the
droplet's parameters, three different scenarios have been observed:
(i) the droplet shrinks until it disappears;
(ii) the droplet grows toward stationary radius where the influx is balanced
by the efflux across the
interface, and therefore it coexists with the surrounding (stable state);
(iii) the droplet undergoes shape instability where any small shape
deformation triggers the elongation along one axis until droplet division.

In the present paper, we investigate the effective model suggested in
\cite{David-Rabea2017} by means of the nonlinear stability theory. In Sec.
\ref{FP}, we briefly describe the mathematical models.
In Sec. \ref{LSA}, we study the instability development of a motionless
one-dimensional interphase boundary. The instability criterion is
obtained, and the equation governing the boundary evolution is derived in
the limit of long waves. In Sec. \ref{MFA}, the analysis is done for a
moving boundary. In Sec. \ref{D2df}, the general case of a
two-dimensional moving interface is considered.

\section{FORMULATION OF THE PROBLEM}\label{FP}

In the present section, we briefly describe the models formulated in the
supplementary
information (SI) of Ref. \cite{David-Rabea2017}.

\subsection{The continuum model }
Let us consider a segregated binary solution with component $B$ dissolved in solvent $A$. The field $u$ describes the concentration of droplet material $B$ both inside and outside the droplet, therefore the phase with a high equilibrium concentration of $B$, $u_-^{(0)}$, forms droplets in the ocean of the phase with a low
concentration, $u_+^{(0)}$. The free energy density function has a double well shape,
\begin{equation*}\label{}
 f(u) = \frac{b}{2(\Delta u)^2 } \left(u-u_-^{(0)} \right)^2 \left(u-u_+^{(0)} \right)^2
\end{equation*}
where the positive parameter $b$ characterizes the molecular interactions and entropic contributions, and
$\Delta u = u_-^{(0)} - u_+^{(0)}>0$.

 The state of the system is described by the free energy functional
\begin{equation*}\label{}
 F[u] = \int_V \left( f(u) + \frac{\kappa}{2} |\nabla u|^2   \right) \di \upsilon,
\end{equation*}
where $V$ is the volume of the system,  $\di \upsilon$ is a volume element, and $\kappa$ is a coefficient,
which determines the surface tension
and the interface width \cite{CahnHilliard58}. Consequently, the chemical potential is given by the variational derivative,
\begin{eqnarray*}\label{}
 && \mu = \frac{\delta F}{\delta u} =\\
 && \frac{b}{|\Delta u|^2}\left( u- u_-^{(0)} \right) \left( u- u_+^{(0)} \right)\left( 2u- u_-^{(0)}- u_+^{(0)} \right) - \kappa \nabla^2 u .\nonumber
\end{eqnarray*}

\begin{figure}
  \centering
  \includegraphics[scale=0.3]{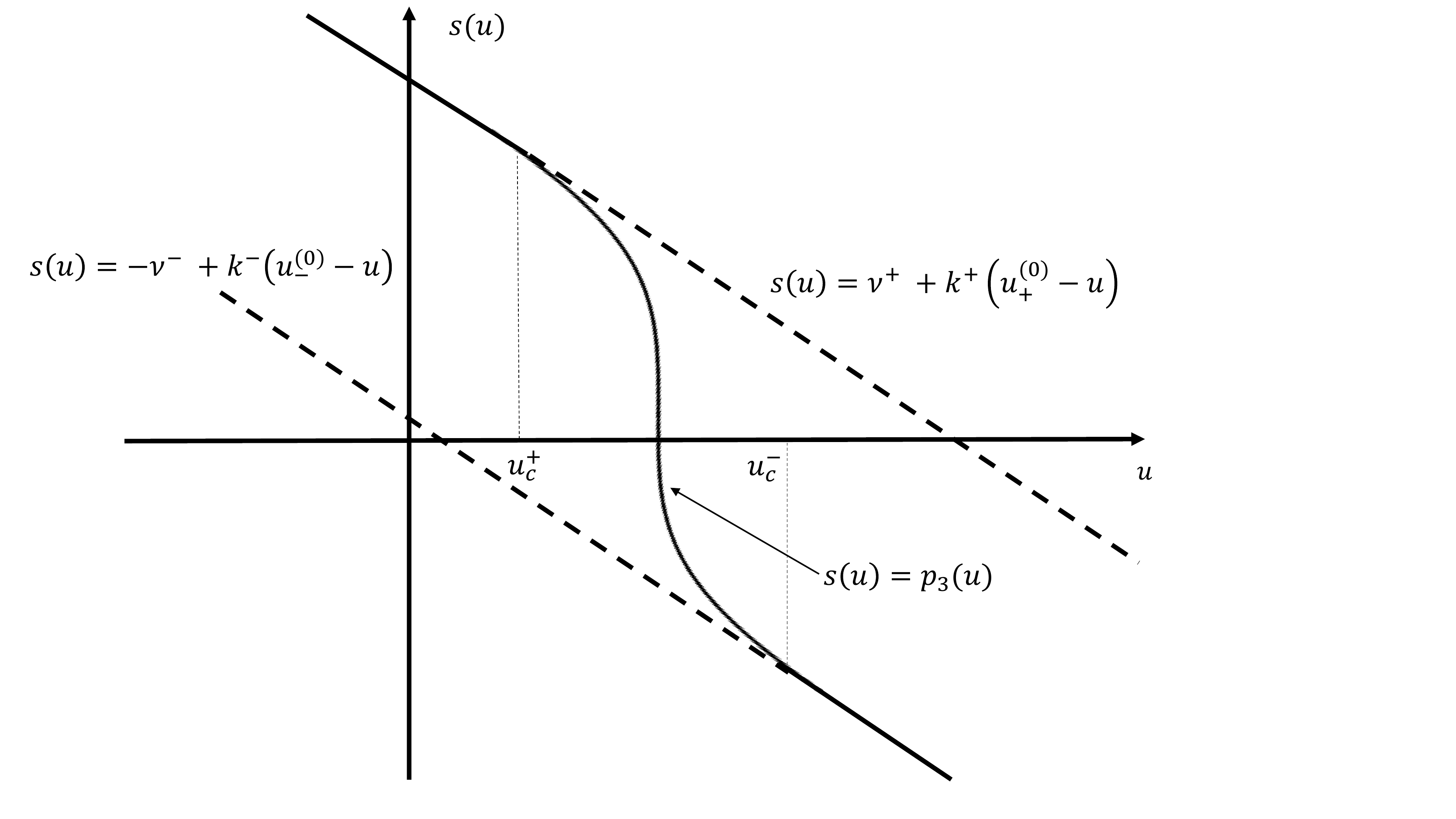}
  \caption{Plot of the source function $s(u)$ in (\ref{sc}) that behaves as cubic polynomial when
$u_c^- < u < u_c^+$ and linear otherwise.  }\label{s(u)}
\end{figure}

The component $B$ is subject to diffusion and chemical reactions. Hence, the material concentration field
dynamics is governed by the reaction diffusion equation,
 \begin{equation}\label{rd}
  u_t = \nabla\cdot( m(u) \nabla \mu) + s(u),
\end{equation}
where $m(u)$ is a mobility coefficient of the component $B$, and the reaction function $s(u)$
is designed
to be linear in the phases outside and inside the droplet, and smoothly interpolated by a cubic polynomial $p_3(u)$
in some interval $u_c^- < u < u_c^+$, where $u_c^{\pm}$ are some characteristic concentrations,
$u_c^-<u_-^{(0)}$, $u_c^+>u_+^{(0)}$, see Fig. \ref{s(u)},
\begin{equation}\label{sc}
s(u) =
\begin{cases}
\nu^+ + k^+ ( u_+^{(0)} -u ) , & u\leq u_c^+,  \\
p_3 (u) , & u_c^+ \leq u \leq u_c^-,  \\
-\nu^- + k^- ( u_-^{(0)} -u ) , & u_c^- \leq  u,
\end{cases}
\end{equation}
here $k^\pm$ are the reaction rate outside and inside the droplet respectively, and similarly $\nu^\pm$ are the
reaction fluxes at equilibrium concentration. We do not present here the explicit expression for
$p_3(u)$, but we notice that the polynomial $p_3(u)$ is uniquely determined.

\subsection{The effective sharp-interface model}\label{Efm}

The model described above determines a diffuse boundary between the phases; its width is
$\delta\sim(\kappa/b)^{1/2}$. Below we assume that the width of the transition layer $\delta$ is small with
respect to any other characteristic scales of the problem, and apply the sharp interface limit. Within that
limit, two boundary conditions are applied at the sharp interface. The first one is the continuity of the
chemical potential across the interface,
 \begin{equation}\label{B1}
\mu(u^-) = \mu(u^+),
\end{equation}
and the second one is the Laplace pressure jump across the interface,
   \begin{equation}\label{B2}
(u^- - u^+) \mu(u^-) + f(u^+) - f(u^-) = 2\sigma H,
\end{equation}
where $u^-$ and $u^+$ are the concentration values at the interface
inside and outside the droplet, respectively, $\sigma=(\Delta u)^2\sqrt{\kappa b}/6$ is the
surface tension or the free
energy per unit of area of the interface, and $H$ is the
local mean curvature of the interface,
   \begin{equation}\label{B3}
H=\frac{1}{2}(\kappa_1+\kappa_2),
\end{equation}
where $\kappa_1$ and $\kappa_2$ are principal curvatures of the interface.
 Equations (\ref{B1}) and (\ref{B2})
are the coexistence conditions of the inside and outside phases that are separated by the droplet interface \cite{DavidZwicker2013}, \cite{WeberDavid2018}.

The next step of the model simplification is the linearization of equations (\ref{rd}), (\ref{sc}). The source
function (\ref{sc}) is approximated by a piecewise linear function,
\begin{equation*}\label{se}
s^L (u) =
\begin{cases}
\nu^+ + k^+ ( u_+^{(0)} -u ) , & \text{outside the droplet} ,  \\
-\nu^- + k^- ( u_-^{(0)} -u ) , & \text{inside the droplet}.
\end{cases}
\end{equation*}
Later on, we will refer the signs ($+$) and ($-$) for the values outside and inside the droplet, respectively.

Linearizing equation (\ref{rd}) around the values $u_-^{(0)}$ and $u_+^{(0)}$ inside and outside the droplet
respectively yields
the linear equations,
\begin{equation}\label{rdbi}
  \p_t u^\pm = D^{\pm}\nabla^2 u^\pm  -m^\pm\kappa\nabla^4 u^\pm + s^L (u),
\end{equation}
where $u^+$ and $u^-$ are the concentration of the droplet material $B$ outside and inside the
droplet respectively,
$D^\pm$ and $m^{\pm}$ are constant coefficients.
Below we omit the bi-harmonic terms $m^\pm\kappa\nabla^4 u^\pm$ in equations (\ref{rdbi}),
following the estimates presented in Ref. \cite{David-Rabea2017}.

In the small surface tension limit, $\sigma \ll \Delta u /(H\beta)$, where $\beta=2/(b\Delta u)$
is
the coefficient that describes the Laplace pressure effect on the interface's concentration, the equilibrium conditions (\ref{B1}) and (\ref{B2}) are approximated as follows,
\begin{eqnarray*}
 && u^- \sim u_-^{(0)} + \beta\sigma H,\\
 && u^+ \sim u_+^{(0)} + \beta\sigma H.
\end{eqnarray*}
The interface dynamics is governed by
\begin{equation}
  v_n =\hat{\textbf{n}} \cdot \frac{\textbf{j}^- - \textbf{j}^+}{u^- - u^+},  \label{gov4}
\end{equation}
where $v_n$ is the normal velocity of the interface, $\hat{\textbf{n}}$ is a unit vector  normal to the interface, and  $\textbf{j}^\pm = -D^\pm\nabla u^\pm $ are the diffusion fluxes outside and inside the droplet, respectively. Recall again that $u^\pm$ in (\ref{gov4}) are the concentrations evaluated outside and inside the droplet interface, respectively.

\subsection{Formulation of the nondimensional local problem }
\begin{figure}
  \centering
  \includegraphics[scale=0.3]{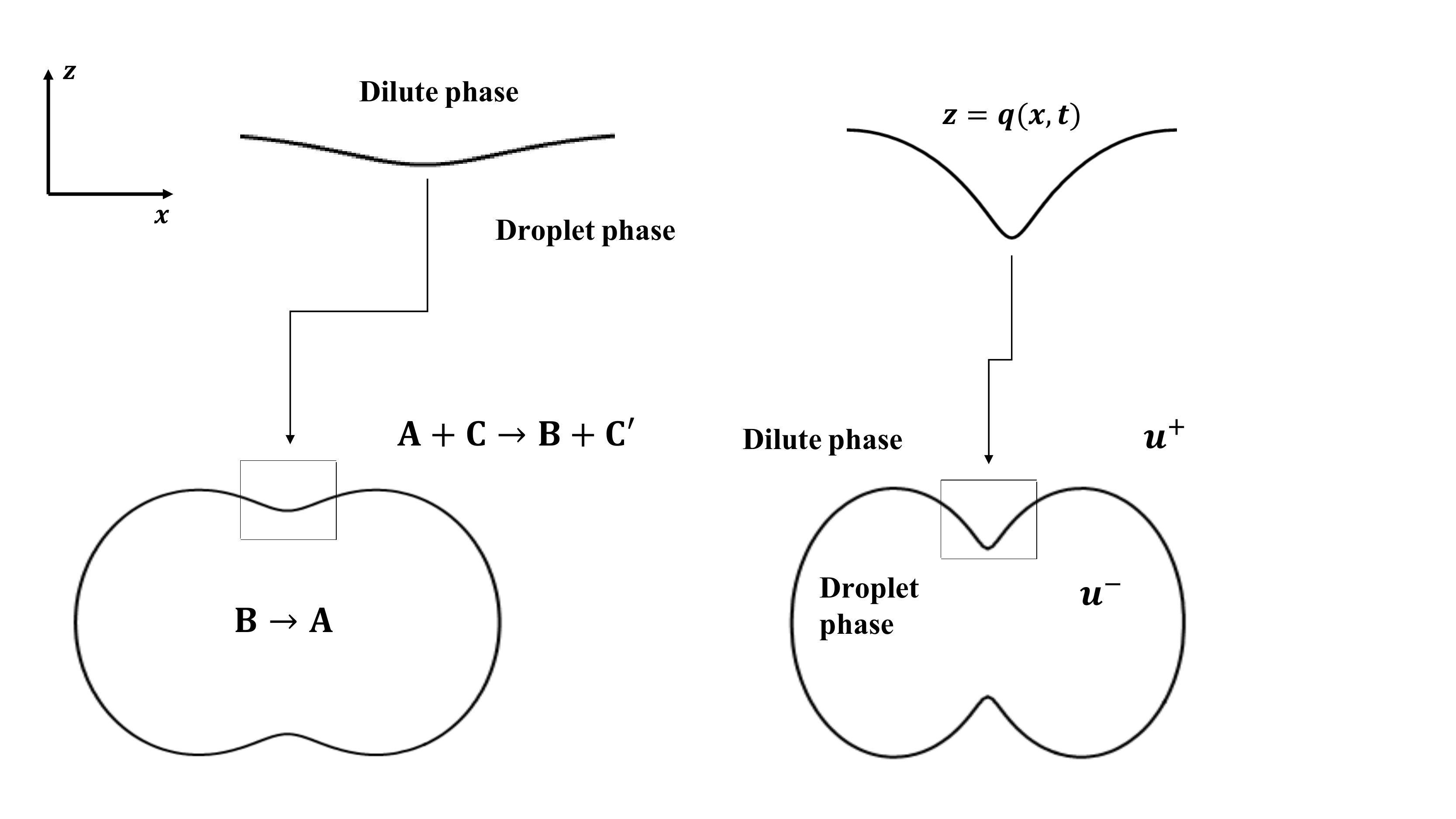}
  \caption{A schematic description of the droplet division dynamics while concentrating (zooming in)
on the region where the division occurs. Inside the droplet, material $B$ is spontaneously
transformed into soluble dilute material $A$ that leaves the droplet. Outside the droplet,
material $A$ is transformed into material $B$, while chemical fuel $C$ is transformed into
product $C'$. Finally,
material $B$ diffuses inside the droplet, thus completing the reaction cycle. }\label{division}
\end{figure}

Both the continuum model and the effective model of the previous sections were studied numerically in the
framework of a spherical droplet in a spherical coordinate system centered at the droplet center (SI \cite{David-Rabea2017}). In
this paper, we concentrate on the region where the division starts, see Fig. \ref{division}. In that region,
we formulate the effective model in the infinite space using Cartesian coordinates $(x,y,z)$, i.e.,
the droplet dynamics is reduced to the dynamics of a distorted {\em flat} interface that could be
described by the function $z=q(x,y,t)$, (see Fig. \ref{division} for the one dimensional description). The goal of
the next sections is to understand the dynamics of such interface using only analytical tools both in
the linear and nonlinear regime.

We begin with the formulation of the effective model in a nondimensional form. We introduce the
concentration, length, and time scales, respectively,
\begin{equation*}
  \Delta u = u_-^{(0)} - u_+^{(0)}, \quad L^+ = \sqrt{\frac{D^+}{k^+}}, \quad  T^+ = \frac{1}{k^+};
\end{equation*}
then we define the scaled variables and fields,
\begin{eqnarray*}
&&(x,y,z) = L^+ (x^*,y^*,z^*), \quad t = T^+ t^*,\\
&& u(x,y,z,t) = (\Delta u) \ u^{**}(L^+ x^*, L^+ y^*, L^+ z^*, T^+ t^* ) = \nonumber\\
&& (\Delta u) \  u^* ( x^*, y^*, z^*, t^* ),\\
&& q(x,y,t) = L^+ q^{**}(L^+ x^* , L^+y^*, T^+ t^* ) = L^+ q^*( x^*, y^*, t^* ),\\
&& H(x,y,t) = (L^+)^{-1} H^{**}( L^+ x^*, L^+y^*, T^+ t^* ) =\\
&& (L^+)^{-1} H^*( x^* , y^*, t^* ).
\end{eqnarray*}
As a result, we obtain six nondimensional parameters of the system,
\begin{eqnarray}
 && N^\pm = \frac{\nu^\pm}{k^+ \Delta u },\quad U^- = \frac{u_- ^{(0)}}{\Delta u},\quad B =
\frac{\beta \sigma}{2L^+ \Delta
u},\nonumber\\
 && D = \frac{D^-}{D^+},\quad K = \frac{k^-}{k^+}.\label{nondp}
\end{eqnarray}
Denote $U^+ =\frac{u_+ ^{(0)}}{\Delta u} = U^- -1 $ and drop the stars, then the local effective model of subsection \ref{Efm} takes the form
\begin{subequations}
\begin{eqnarray}
&&\p_t u^+ = \nabla^2 u^+ - u^+ + U^+ + N^+  , \  z>q(x,y,t),\label{ngov1}\\
&& \p_t u^- = \nonumber\\
&& D \nabla^2 u^- -K u^-  +K U^- -N^- , \quad  z<q(x,y,t),\label{ngov2}\\
&& u^{\pm} = U^\pm  + 2B H(x,y,t); \quad z=q(x,y,t),\label{ngov3}
\end{eqnarray}
\end{subequations}
the nondimensional form of the front dynamics equation is
\begin{eqnarray}
  &&\left( u^- - u^+ \right)\p_t q = \left( -\p_x q, -\p_y q , 1 \right)\cdot\left( \nabla u^+ - D \nabla u^- \right),\nonumber\\
  &&  z=q(x,y,t). \label{ngov4}
\end{eqnarray}

\section{STABILITY ANALYSIS FOR STATIONARY ONE DIMENSIONAL INTERFACE}\label{LSA}
\subsection{Stationary state of the flat interface}\label{base}
First, let us consider the flat {\em stationary} interface $q(x,t)=\bar{q}$; without losing
of generality, we may assume that $\bar{q} =0$. The local curvature vanishes in this case, $H=0$. Also, we assume that the
concentration field is translationally invariant along the $x$ and $y$ axes, i.e., $ u = \bar{u}(z) $. As a result, equations (\ref{ngov1}-\ref{ngov3}) take the form
\begin{subequations}
\begin{eqnarray}
&& \p_z^2 \bar{u}^+ - \bar{u}^+ + U^+ + N^+  =0 , \quad z>0,\label{sngov1}\\
&&  D \p_z^2 \bar{u}^- -K \bar{u}^-  + K U^- -N^- =0 , \quad z<0,\label{sngov2}\\
&& \bar{u}^{\pm} = U^\pm   , \quad z=0;\label{sngov3}
\end{eqnarray}
\end{subequations}
also, we apply the regularity condition
$$|u^\pm (\pm\infty)| < \infty.$$
Equations (\ref{sngov1}-\ref{sngov3}) have the solutions,
\begin{subequations}
\begin{eqnarray}
&& \bar{u}^+ =  U^+ + N^+ \left(1 -  \ex^{ - z}\right),\label{b+}\\
&& \bar{u}^- =U^- - \frac{N^-}{K} \left( 1 - \ex^{z\sqrt{K / D} } \right). \label{b-}
\end{eqnarray}
\end{subequations}

Considering the front dynamics in equations (\ref{gov4}) and (\ref{ngov4}), we conclude that the flat front is
{\em motionless}, $v_n = 0$, only if the influx is balanced by efflux  across the flat interface $q=0$,
\begin{eqnarray}\label{fluxb}
 && \hat{\textbf{n}}\cdot \textbf{j}_- = \hat{\textbf{n}}\cdot\textbf{j}_+ \Leftrightarrow  \p_z \bar{u}^+ = D\p_z \bar{u}^- \Leftrightarrow \nonumber \\
 && N^+  =  N^- \sqrt{\frac{D}{K}}.
 \end{eqnarray}
Equation (\ref{fluxb}) gives us the condition on the system parameters to achieve the stationary state, i.e.,
the coexistence condition for two phases separated by the flat motionless interface.

 In view of equation (\ref{b+}), we find that
 \begin{equation*}\label{}
   N^+ =  \bar{u}^+ (z=\infty) -  \bar{u}^+ (z=0),
 \end{equation*}
therefore, $N^+$ determines the excess concentration of the droplet material far from the droplet. Therefore,
 we call the parameter $N^+$, which will play a central role in the
next sections, the supersaturation parameter.

\subsection{Dispersion relation }
For investigating the stability of the flat front $q=0$ and the stationary solutions (\ref{b+}) and (\ref{b-}),
we introduce the disturbances,
\begin{subequations}
 \begin{eqnarray}
&& u = \bar{u} + \hat{u}, \quad \bar{u} \gg \hat{u}, \label{ud}\\
&& q = \bar{q} + \hat{q}, \quad \bar{q} \gg \hat{q}.\label{qd}
\end{eqnarray}
\end{subequations}
Because of the rotational invariance of the problem, it is sufficient to consider only the two-dimensional disturbances, assuming that
all the variables do not depend on $y$. According to (\ref{B3}), the mean curvature is
\begin{equation*}\label{}
   H(q) =\frac{-\p_x^2 q}{2\left(1+ (\p_x q)^2 \right)^{3/2}} =-\frac{1}{2} \p_x^2 \hat{q}+...
 \end{equation*}
 Substituting the perturbed solution (\ref{ud}),(\ref{qd}), into the system (\ref{ngov1})-(\ref{ngov4}), and
neglecting nonlinear terms, we obtain the linearized problem for disturbances:
\begin{subequations}
 \begin{eqnarray}
&&\p_t \hat{u}^+ = \nabla^2 \hat{u}^+ - \hat{u}^+   , \quad z>0 ,\label{pngov1}\\
&& \p_t \hat{u}^- = D \nabla^2 \hat{u}^- -K \hat{u}^-    , \quad z<0 ,\label{pngov2}\\
&& \hat{u}^{\pm} = - B \p_x^2 \hat{q} - \hat{q} \p_y \bar{u}^\pm , \quad z=0, \label{pngov3}\\
&& \lim_{z\rightarrow\pm\infty} \hat{u}^\pm =0, \label{pngov4}
\end{eqnarray}
\end{subequations}
in addition, we get the interface disturbance equation,
\begin{equation}\label{}
  \p_t \hat{q} = \p_z \hat{u}^+ - D\p_z \hat{u}^- + \left(\p_z^2 \bar{u}^+ - D\p_z^2 \bar{u}^- \right)\hat{q}, \quad z=0. \label{pngov5}
\end{equation}
Let us introduce the normal modes,
\begin{equation}\label{nmode}
 \hat{u}^\pm (x,z,t) = A^\pm (z) \ex^{i\omega x + \sigma t}, \quad \hat{q}(x,t) = Q(\omega,\sigma) \ex^{i\omega x + \sigma t},
\end{equation}
where $\sigma$ is the growth rate, and $\omega$ is the wave number of the disturbance. When substituting
(\ref{nmode}) into the linear system (\ref{pngov1})-(\ref{pngov4}), one finds the solutions,
\begin{subequations}
 \begin{eqnarray}
&&\hat{u}^+ = -Q\left(-\omega^2 B + N^+ \right)\times \label{hu+} \\
&& \exp\left( -z\sqrt{\sigma+\omega^2 +1} + i\omega x + \sigma t\right), \nonumber \\
&&\hat{u}^- = -Q\left(-\omega^2 B + \frac{N^+}{D} \right) \times \label{hu-} \\
&& \exp\left(z\sqrt{\frac{\sigma}{D}+\omega^2 +\frac{K}{D}} + i\omega x + \sigma t\right). \nonumber
\end{eqnarray}
\end{subequations}
Putting these expressions (\ref{hu+}), (\ref{hu-}) into equation (\ref{pngov5}) yields the following dispersion
relation,
\begin{eqnarray}\label{dispr}
 && \sigma = \left(-\omega^2 B + N^+ \right) \sqrt{\sigma + \omega^2 +1} + \\
   && \left(-\omega^2 DB + N^+ \right) \sqrt{\frac{\sigma}{D} + \omega^2 +\frac{K}{D}} - N^+ \left( 1+ \sqrt{\frac{K}{D}}\right), \nonumber
\end{eqnarray}

Note that $\sigma=\omega=0$ is a solution for (\ref{dispr}), which corresponds to a homogeneous shift of the
front. The existence of a neutral mode at $\omega=0$ creates a possibility of a longwave instability at small
nonzero $\omega$. Indeed, assume $\sigma=S\omega^2+o(\omega^2), \quad \omega^2\ll 1$. At the leading order,
equation (\ref{dispr}) yields:
\begin{eqnarray*}
&&S\left[1-\frac{N^+}{2}\left(1+\frac{1}{\sqrt{KD}}\right)\right]= \frac{N^+}{2}\left(1+\sqrt{\frac{D}{K}}\right)\\
&& -B(1+\sqrt{KD}).
\end{eqnarray*}
Thus we find that if
\begin{equation}\label{threshold}
2B\frac{1+\sqrt{KD}}{1+\sqrt{D/K}}< N^+ < \frac{2}{1+1/\sqrt{KD}},
\end{equation}
then $S>0$, which corresponds to a monotonic longwave instability.

\subsection{Longwave nonlinear theory}\label{LWNT}

In the present section, we perform a detailed derivation of the closed longwave nonlinear
equation governing the evolution of the surface
deformation in the case, where all the functions do not depend on $y$. The general case is considered in Section \ref{D2df}.

Recall that the effective model (\ref{ngov1}-\ref{ngov4}) governing the interface dynamics is actually
nonlinear due to the curvature effect.
In this section, we derive the weakly nonlinear evolution equation governing the nonlinear development of
longwave instability. We follow the approach of Sivashinsky
\cite{Sivashinsky1983} in the limit of the small supersaturation number $N^+ \ll 1$.

First let us perform the shift transformations $$u^+ \leftarrow u^+ +U^+ +N^+, \quad u^- \leftarrow u^- + U^- - N^- /K$$ on equations (\ref{ngov1})-(\ref{ngov4}) to obtain the system,
\begin{subequations}
 \begin{eqnarray}
&& u_t^+ = u_{xx}^+ + u_{zz}^+ - u^+, \quad z>q(x,t),\label{shf1}\\
&& u^+ = -N^+ + 2B H(x,t), \quad z=q(x,t),\label{shf2}\\
&& u_t^- = u_{xx}^- + u_{zz}^- -K u^-, \quad z<q(x,t),\label{shf3}\\
&& u^- = \frac{N^-}{K} + 2B H(x,t), \quad z=q(x,t),\label{shf4}\\
&& \left(u^- - u^+ -\frac{N^-}{K} -N^+ +1 \right)q_t = \label{shf5}\\
&& u_z^+ - D u_z^- -q_x \left(u_x^+ - D u_x^- \right), \quad z=q(x,t). \nonumber
\end{eqnarray}
\end{subequations}
Then we introduce the curvilinear coordinates,
\begin{equation*}\label{}
  \tilde{z} = z-q(x,t),\quad \tilde{x}=x,\quad \tilde{t}=t,
\end{equation*}
and define,
\begin{eqnarray*}\label{}
  && q(x,t) = q(\tilde{x},\tilde{t}) = \tilde{q}(\tilde{x},\tilde{t}), \quad H(q)=\tilde{H}(\tilde{q})\\
  && u(x,z,t) = u(\tilde{x},\tilde{z} + \tilde{q}(\tilde{x},\tilde{t}) ,\tilde{t}) = \tilde{u}(\tilde{x},\tilde{z},\tilde{t}).
\end{eqnarray*}
Applying the chain rule and dropping the tildes, we transform equations (\ref{shf1})-(\ref{shf5}) to the form
 \begin{subequations}
 \begin{eqnarray}
&& u_t^+ - q_t u_z^+ =\label{crv1} \\
&& u_{zz}^+ + u_{xx}^+ + q_x^2 u_{zz}^+ -2q_x u_{xz}^+ -q_{xx}u_z^+ - u^+, \quad z>0, \nonumber\\
&& u^+ = -N^+ + 2B H(x,t), \quad z=0,\label{crv2}\\
&& u_t^- - q_t u_z^- =\label{crv3} \\
&& D( u_{zz}^- + u_{xx}^- + q_x^2 u_{zz}^- -2q_x u_{xz}^- -q_{xx}u_z^- ) - K u^-, \ \  z<0, \nonumber\\
&& u^+ = \frac{N^-}{K} + 2B H(x,t), \quad z=0,\label{crv4}\\
&& \left(u^- - u^+ -\frac{N^-}{K} -N^+ +1 \right)q_t =\label{crv5}\\
&& \left(q_x^2 +1\right)\left( u_z^+ - D u_z^- \right) -q_x \left(u_x^+ - D u_x^- \right) , \quad z=0. \nonumber
\end{eqnarray}
\end{subequations}

In order to obtain a closed amplitude equation for the surface deformation, we assume the following scaling of
the system variables and parameters:
 \begin{eqnarray}
 && \tau = \ep^6 t, \quad \xi = \ep x, \quad \zeta =z,\nonumber \\
 && N^\pm = \ep^2 \left( \Lambda_c^\pm + \ep^2 \Lambda_2^\pm \right), \quad B = \ep^2 \Phi ,\label{scal}
 \end{eqnarray}
where all the Greek upper case letters denote quantities $O(1)$. Note that according to (\ref{fluxb})
$$\Lambda_c^+=\Lambda_c^-\sqrt{\frac{D}{K}},\;\Lambda_2^+=\Lambda_2^-\sqrt{\frac{D}{K}}.$$
The front dynamics is slow in time and large-scale in space.
Motivated by the stationary solutions (\ref{b+})-(\ref{b-}), we assume the following asymptotic expansions of the
fields,
\begin{subequations}
\begin{eqnarray}
&& u = \ep^2 u_2(\xi,\zeta,\tau) + \ep^3 u_3 + \ep^4 u_4 + ..., \label{ue}\\
&& q= q_0(\xi,\tau) + \ep q_1 +.... \label{qe}
\end{eqnarray}
\end{subequations}
Next, we substitute (\ref{ue}) and (\ref{qe}) in equations (\ref{crv1})-(\ref{crv5}), and collect the terms of the same order.

At the leading order $O(\ep^2)$, we obtain the system of equations that is equivalent to the base solutions that were considered in subsection \ref{base},
\begin{eqnarray}
&& u_{2\zeta\zeta}^+ - u_2^+ =0 ,\quad \zeta>0,\nonumber\\
&& u_2^+ = -\Lambda_c^+, \quad\zeta=0,\nonumber\\
&& u_{2\zeta\zeta}^- - \frac{K}{D} u_2^- =0 ,\quad \zeta<0,\nonumber\\
&& u_2^- = \frac{\Lambda_c^-}{K}, \quad \zeta=0,\nonumber\\
&& u_{2\zeta}^+ - D u_{2\zeta}^- =0, \quad \zeta=0. \label{fb2}
\end{eqnarray}
The solutions are,
\begin{equation*}\label{u2}
  u_2^+ = -\Lambda_c^+ \ex^{-\zeta}, \quad u_2^- = \frac{\Lambda_c^-}{K} \ex^{\zeta\sqrt{K/D}},
\end{equation*}
then indeed, equation (\ref{fb2}) is equivalent to the flux balance condition (\ref{fluxb}). At the next order $O(\ep^3)$, the equations are homogenous therefore $u_3^\pm\equiv0$.

At order $O(\ep^4)$, we have the system,
\begin{eqnarray}
&& u_{4\zeta\zeta}^+ - u_4^+ = A_4^+ \ex^{-\zeta} ,\quad \zeta>0, \nonumber\\
&& u_4^+ = -\Lambda_2^+ -\Phi q_{0\xi\xi} , \quad \zeta=0,\nonumber\\
&& u_{4\zeta\zeta}^- - \frac{K}{D} u_4^- = A_4^- \ex^{\zeta\sqrt{K/D}} ,\quad \zeta<0,\nonumber\\
&& u_4^- =  \frac{\Lambda_2^+}{K} -\Phi q_{0\xi\xi} , \quad \zeta=0,\nonumber\\
&&  u_{4\zeta}^+ - D u_{4\zeta}^- =0, \quad \zeta=0, \label{fb4},
\end{eqnarray}
where
\begin{eqnarray*}
&& A_4^+  = \Lambda_c^+ \left( q_{0\xi}^2 + q_{0\xi\xi} \right) \\
&& A_4^-  = -\frac{\Lambda_c^-}{DK} \left( K q_{0\xi}^2  - \sqrt{DK} q_{0\xi\xi} \right),
\end{eqnarray*}
hence one can calculate the solutions,
\begin{eqnarray*}
 &&  u_4^+ = - \left( \Lambda_2^+ +\Phi q_{0\xi\xi} + \frac{A_4^+}{2}\zeta \right)\ex^{-\zeta}, \nonumber \\
 &&  u_4^- = \left( \frac{\Lambda_2^+}{K} - \Phi q_{0\xi\xi} + \frac{A_4^-}{2}\sqrt{\frac{D}{K}} \zeta \right) \ex^{\zeta\sqrt{K/D}}.\label{u4}
\end{eqnarray*}
As a result, equation (\ref{fb4}) yields the equation,
\begin{equation*}\label{}
 \left[-\frac{\Lambda_c^+}{2} \left(1+ \sqrt{\frac{D}{K}}\right) + \Phi (1+ \sqrt{DK}) \right]q_{0\xi\xi} = 0,
 \end{equation*}
Which determines the instability threshold,
\begin{equation}\label{reco}
 \Lambda_c^+ = \frac{2\Phi(1+\sqrt{DK})}{1+\sqrt{D/K}}
 \end{equation}
Note that (\ref{reco}) coincides with (\ref{threshold}).

At order $O(\ep^5)$, we have the system,
\begin{eqnarray}
&& u_{5\zeta\zeta}^+ - u_5^+ = A_5^+ \ex^{-\zeta} ,\quad \zeta>0, \nonumber\\
&& u_{5\zeta\zeta}^- - \frac{K}{D} u_5^- = A_5^- \ex^{\zeta\sqrt{K/D}} ,\quad \zeta<0,\nonumber\\
&& u_5^\pm =   -\Phi q_{1\xi\xi} , \quad \zeta=0,\nonumber\\
&&  u_{5\zeta}^+ - D u_{5\zeta}^- =0, \quad \zeta=0, \label{fb5},
\end{eqnarray}
where
\begin{eqnarray*}
&& A_5^+  = \Lambda_c^+ \left( 2q_{0\xi}q_{1\xi}  + q_{1\xi\xi} \right) \\
&& A_5^-  = \frac{\Lambda_c^-}{DK} \left( -2K q_{0\xi} q_{1\xi}  + \sqrt{DK} q_{1\xi\xi} \right),
\end{eqnarray*}
hence one can calculate the solutions,
\begin{eqnarray*}
 &&  u_5^+ = - \left( \Phi q_{1\xi\xi} + \frac{A_5^+}{2}\zeta \right)\ex^{-\zeta}, \nonumber \\
 &&  u_5^- = \left( -\Phi q_{1\xi\xi} + \frac{A_5^-}{2}\sqrt{\frac{D}{K}} \zeta \right) \ex^{\zeta\sqrt{K/D}},\label{u5}
\end{eqnarray*}
then equation (\ref{fb5}) is satisfied due to condition (\ref{reco}).

At the next order, $O(\ep^6)$, we will obtain a closed equation
governing the interface dynamics $q_0(\xi,\tau)$. It holds that,
 \begin{eqnarray}
&& u_{6\zeta\zeta}^+ - u_6^+ = (E^+ + \zeta F^+) \ex^{-\zeta} ,\quad \zeta>0, \nonumber\\
&& u_{6\zeta\zeta}^- - \frac{K}{D} u_6^- = (E^- + \zeta F^-) \ex^{\zeta\sqrt{K/D}} ,\quad \zeta<0,\nonumber\\
&& u_6^\pm =  \Phi\left( \frac{3}{2} q_{0\xi}^2 q_{0\xi\xi} - q_{1\xi\xi} \right) , \quad \zeta=0,\nonumber\\
&& q_{0\tau}= u_{6\zeta}^+ - D u_{6\zeta}^- -q_{0\xi} \left( u_{4\xi}^+ - D u_{4\xi}^- \right)  , \quad \zeta=0,
\label{fb6}
\end{eqnarray}
where $E^\pm$, and $F^\pm$ are tedious expression that include $q_0, q_1, q_2, u_4^\pm$, and their derivatives, see Appendix \ref{App:AppendixA}. One can calculate,
\begin{eqnarray}
 &&  u_6^+ = \left[ u_6^+(\zeta=0)  - \left(\frac{ E^+}{2} + \frac{F^+}{4} \right)\zeta -\frac{F^+}{4}\zeta^2 \right] \ex^{-\zeta}, \nonumber \\
 &&  u_6^- = \Big[ u_6^-(\zeta=0)  + \left(\frac{ E^-}{2}\sqrt{\frac{D}{K}} - \frac{D F^-}{4K} \right)\zeta \nonumber  \\
 &&  +\frac{F^+}{4}\sqrt{\frac{D}{K}} \zeta^2 \Big]  \ex^{\zeta\sqrt{K/D}},\nonumber
\end{eqnarray}
therefore equation (\ref{fb6}) yields the following amplitude equation for the interface at
leading order,
\begin{equation}\label{q0}
\p_\tau q_0 = -\alpha_1 \p_\xi^2 q_0 - \beta_1 \p_\xi^4 q_0 + \gamma_1  \left( \p_\xi^2 q_0\right)^2,
\end{equation}
where,
\begin{eqnarray}
 && \alpha_1  =\frac{1}{2}\Lambda_2^+ \left(1+ \frac{D}{K}\right)\nonumber\\
 && \beta_1 = \frac{1}{2} \left[ \Phi\left(1+D\sqrt{\frac{D}{K}}\right)  +   \frac{\Lambda_c^+}{4}\left(1+ \left[\frac{D}{K}\right]^{3/2} \right) \right] \nonumber\\
 && \gamma_1 = \frac{\Phi}{4} \left[ 3(D-1) + \sqrt{\frac{D}{K}}(1-K) \right].
\end{eqnarray}

Let us emphasize that the nonlinear term in the obtained amplitude equation is different from the
$(\partial_{\xi}q_0)^2$ term characteristic for the {\em Kuramoto-Sivashinsky equation}, which is typical for
instabilities of reaction fronts and phase transition fronts. The reason is that we consider here the longwave
instability of a {\em motionless} reaction front (see (\ref{fluxb})), while the above-mentioned nonlinear term
is related to the nonlinearity of the expression for $v_n$ (see (\ref{gov4}), and (\ref{ngov4})). Note that equation (\ref{q0}) can be
transformed to the {\em Sivashinsky equation},
\begin{equation}\label{Siv}
\partial_{\tau}Q= -\alpha_1 \partial_{\xi}^2 Q- \beta_1 \partial_{\xi}^4 Q + \gamma_1 \partial_{\xi}^2(Q^2),
\end{equation}
where $Q=\partial_{\xi}^2q_0$.

In a contradistinction to the Kuramoto-Sivashinsky equation, which describes the
development of a spatio-temporal chaos \cite{YamadaKuramoto1976},
the temporal evolution of a disturbance governed by the Sivashinsky equation leads to formation of singularity. Below we briefly describe the
results of the paper \cite{BernoffBertozzi1995} that contains a detailed description of that singularity.

Near the singularity point,
$$ \partial_{\xi}^2(Q^2) \gg \partial_{\xi}^2 Q,$$
 therefore at the leading order one
can disregard the first term in the right-hand side of (\ref{Siv}). By applying the transformation
$(\beta_1/\gamma_1) g(\xi,\bar{\tau}) = q_{0\xi\xi}$, and $\bar{\tau} = \beta_1\tau$ on
equation (\ref{q0}), one obtains the universal equation
 \begin{equation*}\label{appg}
 g_{\bar{\tau}} +  \p_\xi^4 g - \left( g^2\right)_{\xi\xi} = 0.
\end{equation*}
Its self-similar solution,
\begin{eqnarray*}\label{g}
&& g(\xi,\bar{\tau}) = (\bar{\tau}_c- \bar{\tau})^{-1/2} f(r),\\
&& \quad r = \frac{\xi - \xi_c}{(\bar{\tau}_c - \bar{\tau})^{1/4}}\nonumber
\end{eqnarray*}
describes the formation of a singularity at finite time $\bar{\tau}=\bar{\tau}_c$ in the
point $\xi=\xi_c$. Here
$f$ is a smooth function which satisfies an ordinary differential equation,
\begin{equation*}\label{f}
 \frac{1}{2} \left(f+\frac{1}{2}rf_r \right) = (f^2)_{rr} - \p_r^4 f, \quad -\infty < r <
\infty.
\end{equation*}
For the singularity of $q(\xi,\tau)$, the following relations have been obtained
\cite{BernoffBertozzi1995}:

\begin{subequations}
\begin{eqnarray}
&& q(\xi_c, \tau) \sim -  \frac{\gamma_1C}{\bar{\beta}}\ln(\tau_c - \tau)+ A_0 + O\Big( \sqrt{\tau_c - \tau} \Big),\label{sinq1}\\
&& C\approx 2.01, \nonumber \\
&& q(\xi,\tau) \sim  -  \frac{4\gamma_1C}{\bar{\beta}}\ln|\xi - \xi_c|+ B_0 +\label{sinq2}\\
&& O\left( \sqrt{\tau_c - \tau}, \frac{\tau_c - \tau}{(\xi - \xi_c)^4} \right), \quad (\tau_c - \tau)^{1/4} \ll |\xi-\xi_c|\ll1,\nonumber
\end{eqnarray}
\end{subequations}
where $A_0$, and $B_0$ are constant.

Note that the singularity of the interface distortion propagates towards the droplet
phase (that would correspond to the droplet division) only if $\gamma_1<0$.

\section{MOVING FRONT ANALYSIS}\label{MFA}
If the relation (\ref{fluxb}) is violated, the flat front moves.
In this section we perform the analysis of a moving one dimensional front solution both in the linear and
nonlinear regime.
Therefore we assume a moving front $q(x,t) = v t + h(x,t)$, and consider a traveling solution $u(x,z,t)=\tilde{u}(x,\tilde{z},t)$, where
$\tilde{z}=z-v t$. We drop the tildes then equations (\ref{ngov1}-\ref{ngov4}) take the form,
\begin{eqnarray*}
&&\p_t u^+ = \nabla^2 u^+ +vu_z^+ - u^+ + U^+ + N^+  , \quad z>h(x,t),\label{mngov1}\\
&& \p_t u^- = \nonumber\\
&& D \nabla^2 u^- +vu_z^- -K u^-  +K U^- -N^- , \quad  z<h(x,t),\label{mngov2}\\
&& u^{\pm} = U^\pm  + 2B H(x,t), \quad z=h(x,t),\label{mngov3}\\
&& \left( u^- - u^+ \right) ( v+ h_t) =\nonumber \\
&&\left( - h_x , 1 \right)\cdot\left( \nabla u^+ - D \nabla u^- \right), \quad z=h(x,t). \label{mngov4}
\end{eqnarray*}
A moving flat front solution takes the form,
\begin{eqnarray*}
&& \bar{u}^+ =  U^+ + N^+ \left(1 -  \ex^{z r_- }\right), \quad r_- = \frac{-v-\sqrt{v^2 +4}}{2}  \label{mb+}\\
&& \bar{u}^- =U^- - \frac{N^-}{K} \left( 1 - \ex^{ z r_+  } \right), \quad r_+ = \frac{-v+\sqrt{v^2 +4DK}}{2D}  \label{mb-},
\end{eqnarray*}
while the flux balance (\ref{fluxb}) gives
\begin{equation}\label{mfluxb}
  v+ N^+ r_- + \frac{DN^-}{K} r_+ = 0.
\end{equation}
The dispersion relation (\ref{dispr}) is generalized to
\begin{eqnarray*}
&& \sigma =  m_- ( \omega^2 B + N^+ r_- ) -Dm_+ \left( \omega^2 B -\frac{ N^- r_+}{K} \right) - \nonumber\\
&& N^+ r_-^2 -\frac{DN^-}{K}r_+^2 , \label{dispV} \\
&& m_- = \frac{1}{2}\left( -v - \sqrt{v^2 +4(\sigma+\omega^2 +1)} \right)<0, \\
&& m_+ = \frac{1}{2D}\left( -v + \sqrt{v^2 +4D(\sigma+D\omega^2 +K)} \right)>0.
\end{eqnarray*}
Following the procedure of the nonlinear longwave analysis described in Section
\ref{LWNT},  and
choosing the velocity scaling $v=\ep^4 V$, $V=V_0+\ep^2V_2+\cdots$ in addition to (\ref{scal}),  we obtain the
relations
\begin{eqnarray}
&& \Lambda_c^+ - \sqrt{\frac{D}{K}} \Lambda_c^- =0, \nonumber\\
&& \Lambda_2^+ - \sqrt{\frac{D}{K}} \Lambda_2^- = V_0,\label{V0} \\
&& -\frac{1}{2} \left( \Lambda_c^+ + \frac{D}{K} \Lambda_c^- \right)+ (1+\sqrt{DK})\Phi = 0 \nonumber,
\end{eqnarray}
which are in agreement with (\ref{mfluxb}).
Finally, the moving interface dynamics at the leading order is governed by the following
equation,
\begin{eqnarray}
&& \p_\tau h_0 + \alpha_1 \p_\xi^2 h_0 + \beta_1 \p_\xi^4 h_0 - \frac{V_0}{2} \left( \p_\xi h_0\right)^2  - \gamma_1  \left( \p_\xi^2 h_0\right)^2 =V_2,\nonumber\\
&&  V_2 = \frac{V_0}{2}\left( \Lambda_c^+ + \frac{\Lambda_c^-}{K} \right).\label{V1}
\end{eqnarray}
 We may write $h_0 = V_2 \tau + \psi$, then we obtain the {\em modified Kuramoto-Sivashinsky equation} \cite{BernoffBertozzi1995}
\begin{equation}\label{MKSE}
 \psi_\tau + \alpha_1  \p_\xi^2 \psi + \beta_1 \p_\xi^4 \psi - \frac{V_0}{2} (\p_\xi\psi)^2 - \gamma_1 (\p_\xi^2 \psi)^2 =0,
\end{equation}
which contains two nonlinear terms.

The structure of the obtained equation can be understood in the following way. The normal
velocity $v_n$ of the one-dimensional phase transition boundary is determined by its curvature
$\kappa=2H$. If the curvature is small and slowly depends on the coordinate, the long-wave
expansion for $v_n$ is
\begin{equation}\label{FS}
v_n\sim V-\alpha_1\kappa -\beta_1\partial_s^2\kappa + \gamma_1\kappa^2+\ldots,
\end{equation}
where $V$ is the velocity of the flat boundary, $\partial_s$ is the derivative along the
boundary, $\alpha_1, \beta_1$ and $\gamma_1$ are constant coefficients \cite{FrankelSivashinsky1988}.
For small deformations, expression (\ref{FS}) leads to  (\ref{V1}). The term proportional to
$(\p_\xi\psi)^2$ has a kinematic origin: it is caused by the relation
\begin{equation*}\label{vn}
v_n=\frac{\partial_th}{\sqrt{1+(\partial_xh)^2}}.
\end{equation*}
Therefore, that term, which is characteristic for the standard Kuramoto-Sivashinsky equation, is
ubiquitous in deformational instabilities of moving fronts.
 The term proportional to $(\p_\xi^2\psi)^2$, which is caused by the dependence of the normal
velocity on the curvature, is the same as in (\ref{q0}), and it creates a finite-time
singularity, which corresponds to creation of a caustic for the curve moving according to
(\ref{FS}).

\section{DYNAMICS OF THE TWO DIMENSIONAL INTERFACE}\label{D2df}
The generalization of the derivation presented above to the moving two dimensional interface $q(x,y,t)= vt+h(x,y,t)$
is straightforward.

Expression (\ref{B3}) is replaced with
\begin{equation*}\label{}
 H= -\frac{1}{2} \frac{\nabla^2_{\perp}q + q_{xx}q_y^2 - 2q_x q_y q_{xy} + q_{yy}q_x^2 }{(1+|\nabla_{\perp} q|^2)^{3/2}},
\end{equation*}
where
$$\nabla^2_{\perp}q=q_{xx}+q_{yy},\quad |\nabla_{\perp} q|^2=q_x^2+q_y^2.$$
Using the expansion
\begin{equation*}\label{}
 h(x,y,t) = h_0 + \ep h_1+ \ep^2 h_2 +...,
\end{equation*}
and rescaling the coordinates as $\xi=\ep x$, $\eta=\ep y$,
we obtain the following expansion of the curvature,
\begin{eqnarray*}
 && H = -\frac{\ep^2}{2} \nabla_{\perp}^2 h_0 -\frac{\ep^3}{2} \nabla_{\perp}^2 h_1+ \ep^4 \Big[ -\frac{1}{2} \nabla_{\perp}^2 h_2 + \nonumber\\
 && \frac{3}{4}( h_{0\xi}^2 h_{0\xi\xi} + h_{0\eta}^2 h_{0\eta\eta} ) + \nonumber\\
 && \frac{1}{4}( h_{0\xi}^2h_{0\eta\eta} + h_{0\eta}^2h_{0\xi\xi} ) +h_{0\xi}h_{0\eta}h_{0\xi \eta} \Big]+... .
\end{eqnarray*}
The results of the linear theory of section \ref{LSA} are unchanged, with $\omega$ being the modulus of the wavevector.
The weakly nonlinear analysis gives the following evolutionary equation of the interface at the leading order,
\begin{eqnarray}
 && \p_\tau \psi = -\alpha_1 \nabla_{\perp}^2\psi- \beta_1 \nabla_{\perp}^4\psi + \gamma_2 \left( \nabla_{\perp}^2\psi\right)^2 \nonumber \\
 &&+\frac{V_0}{2} |\nabla_{\perp}\psi|^2 - \delta \left(|\nabla_{\perp}\psi_x|^2| + \nabla_{\perp}\psi_z|^2 \right) \label{2Dh_0},
\end{eqnarray}
where $h_0=V_2\tau+\psi$ and
\[
\delta=\frac{\Phi}{2}(1+\sqrt{DK})\left(1-\sqrt{\frac{D}{K}}\right),\quad \gamma_2=\gamma_1+\delta.
\]
Using the relation
\begin{equation*}
\left(\nabla_{\perp}^2\psi \right)^2 - \left( \left|\nabla_{\perp}\psi_x \right|^2 + \left|\nabla_{\perp}\psi_y \right|^2 \right)  =
 2\left( \psi_{xx}\psi_{yy} - (\psi_{xy})^2 \right),
\end{equation*}
we can rewrite equation (\ref{2Dh_0}) as
\begin{eqnarray}
 && \p_\tau\psi = -\alpha_1 \nabla_{\perp}^2\psi- \beta_1 \nabla_{\perp}^4\psi + \gamma_1\left( \nabla_{\perp}^2\psi\right)^2 \nonumber \\
 && -\frac{V_0}{2} |\nabla_{\perp}\psi|^2+2\delta(\psi_{xx}\psi_{yy}-\psi_{xy}^2).  \label{2Dpsi}
\end{eqnarray}
The evolution equation (\ref{2Dpsi}) corresponds to the longwave limit of the following
expression for the normal velocity \cite{FrankelSivashinsky1988}:
$$v_n\sim V-\alpha_1\kappa-\beta_1\nabla_s^2\kappa+\gamma_1\kappa^2+2\delta G+...,$$
where $\kappa=2H$ and
$$G=\frac{q_{xx}q_{yy}-q_{xy}^2}{(1+|\nabla_{\perp}q|^2)^2}$$
is the Gaussian curvature of the surface.
The development of the singularity for that equation was studied in \cite{BernoffBertozzi1995}.
The one-dimensional case, (\ref{q0}) is recovered.

\section{Conclusion}
We have obtained the condition on the system parameters (\ref{fluxb})
 to achieve the coexistence of two phases separated by the motionless interface. In the case of moving front, an
exact expression of the local interface normal velocity has been obtained (\ref{V0}),(\ref{V1}).
We have carried out the linear stability analysis of the flat interface and found a monotonic
longwave instability. A weakly nonlinear theory near the instability threshold has been
developed. The local interface dynamics is governed by the modified Kuramoto-Sivashinsky equation
that develops a logarithmic singularity after a finite time (\ref{q0}),(\ref{MKSE}),(\ref{2Dh_0}).

\section*{ACKNOWLEDGEMENT}
M. A. H thanks J\"{o}rn Dunkel for introducing him to the inspiring work of the Max Planck Institute group \cite{David-Rabea2017}, and thanks David Zwicker for helpful discussions.

\appendix

\section{} \label{App:AppendixA}
Expressions for $E^\pm$, and $F^\pm$,
\begin{eqnarray}
 && E^+ = \Lambda_c^+ ( q_{1\xi}^2 +2q_{0\xi}q_{2\xi} + q_{2\xi\xi} )+ q_{0\xi\xi}(\Lambda_2^+ + \Phi q_{0\xi\xi} - A_4^+ /2 ) \nonumber \\
 && -q_{0\xi}^2(-\Lambda_2^+ -\Phi q_{0\xi\xi} + A_4^+ ) + 2q_{0\xi} (-A_{4\xi}^+ /2 + \Phi q_{0\xi\xi\xi}) + \Phi \p_\xi^4 q_0. \nonumber \\
 &&F^+ = q_{0\xi\xi} A_4^+ /2 + q_{0\xi}^2 A_4^+ /2 + q_{0\xi}A_{4\xi}^+ + A_{4\xi\xi}^+ /2 .\nonumber \\
  && E^- = -\frac{\Lambda_c^-}{D} \Big( q_{1\xi}^2 +2q_{0\xi}q_{2\xi} - \sqrt{\frac{D}{K}} q_{2\xi\xi} \Big)+ \nonumber\\
 && q_{0\xi\xi}\left( \sqrt{\frac{D}{K}} (\Lambda_2^- /K - \Phi q_{0\xi\xi}) + \sqrt{\frac{D}{K}} A_4^- /2 \right) \nonumber \\
 && -q_{0\xi}^2 \left( \frac{K}{D} (\Lambda_2^- /K -\Phi q_{0\xi\xi} )+ A_4^- \right) + \nonumber\\
  && 2q_{0\xi} \left( \sqrt{\frac{D}{K}}A_{4\xi}^- /2 - \sqrt{\frac{K}{D}} \Phi q_{0\xi\xi\xi} \right) + \Phi \p_\xi^4 q_0. \nonumber \\
 && F^- = q_{0\xi\xi} A_4^- /2 - \sqrt{\frac{K}{D}} q_{0\xi}^2 A_4^- /2 + q_{0\xi}A_{4\xi}^- -\sqrt{\frac{D}{K}} A_{4\xi\xi}^- /2. \nonumber
  \end{eqnarray}

\bibliographystyle{apsrev4-1}
\bibliography{DWEM-7.bbl}{}

\end{document}